# A microscopic quantal self-consistent cranking model for oscillations in spherical nuclei

P Gulshani

NUTECH Services, 3313 Fenwick Crescent, Mississauga, Ontario, Canada L5L 5N1
Tel. #: 905-569-8233; matlap@bell.net

In this article, we transform the previously-derived microscopic rotational-model Schrödinger equation into a form suitable for describing oscillations-coupled-to-intrinsic motion in spherical nuclei. The resulting equation is decomposed into two coupled cranking-type equations, one for the oscillation and another for the intrinsic motion, using a product wavefunction and a constrained variational method. The energy and cranking parameters in the coupled equations are self-consistently determined as functions of the system parameters by the solutions of the coupled equations. This self-consistency makes the two equations time-reversal invariant, unlike the conventional phenomenological cranking models. The self-consistency and time-reversal invariance accept only real solutions to the equations. For the harmonic oscillator mean-field potential, we explicitly determine these solutions and the corresponding eigenvalues, and derive the set of equations that determine self-consistently the parameters. To explore the relative importance of the various model features and approximations, we perform a preliminary scoping calculation of the excitation energy of the first excited $0^+$ states in the light nuclei using a sum rule to determine the oscillation frequency. The preliminary results indicate that, except in the lightest nuclei, the excitation energies are significantly overpredicted in the light nuclei due to the neglect, among other factors, of the deformation degree of freedom. The model derivation presented here serves as guide for eventually developing a corresponding model for the vibrational-rotational motion in deformed nuclei.

1.  **Introduction**

    A nucleus may exhibit excited states that are associated with the motion of a single nucleon or coherent, correlated, or collective modes of the motion of a collection of nucleons. The collective modes are referred to as fission, giant monopole, dipole, quadrupole, octupole, (etc.) resonances (high energy modes), quadrupole-shape rotation and vibration (low energy modes), low-energy breathing mode (including surface and whole-body density oscillations) where the neutrons and protons oscillate in-phase, and whole-body translation. There has been many analytical and experimental studies of the collective nuclear modes of motion.

    In the theoretical studies of any of these modes, one identifies the associated collective variables or co-ordinates and momenta and uses one of the many available methods (such as quantization of classical liquid drop or hydrodynamic models [1,2,3,4,5,6,7,8], sum-rule methods [7,9-16], Tamm-Dancoff, RPA, generator co-ordinate, time-dependent Hartree-Fock method, and equation of motion method [2-6,9,10, 11,17-23], cranking model and its self-consistent constrained Hartree-Fock method [2-6,9,24-55], canonical transformations and algebraic methods [3,4,5,10,56-62], and scaling method [23,63-66]) to formulate the equations of motion for the collective variables. In formulating and solving these equations, assumptions have been made about the nature of the coupling between the collective and the

system's other (intrinsic) motions. For example, it is sometimes assumed that the amplitude of the collective motion is small or the collective motion is slow or adiabatic (i.e., their frequencies or energies are small compared to the intrinsic ones), and hence one can ignore the coupling or treat it as a perturbation.

In the above-mentioned formulations, with the exception of canonical transformation, the nature of the coupling between the collective and intrinsic degrees of freedom is semi-classical and/or phenomenological, and is often treated as a constraint on the intrinsic system as in the cranked Hartree-Fock formulation. Consequently, the coefficient of the coupling operator is a constant parameter, independent of the particle co-ordinates and momenta. This type of coupling results in a theory that is semi-classical, phenomenological, non-self-consistent, and time-reversal non-invariant, leaving one with the task of justifying the assumptions and approximations used for the coupling, and querying their impact on and the accuracy and interpretation of the predicted results.

In this article, we derive a microscopic, quantal, self-consistent cranking model for oscillations in spherical nuclei to quantify the importance and the effects of the self-consistency, time-reversal invariance, and cranking parameters in the conventional phenomenological cranking model. The main goal of this development is to serve as guide for deriving the more involved cranking model for vibrational-rotational motion in deformed nuclei. We generalize the transformation in [62] from the two to three-dimensional space and decompose the Bohr-Mottelson rotational model Hamiltonian developed previously in [61] into the sum of three parts, one describing nuclear radius oscillations, another describing the intrinsic motion, and the third part describing the coupling between these two motions.

An advantage of this model is that it avoids using redundant co-ordinates and/or imposing constraints on the particle co-ordinates (which has complicated utilization of related canonical transformation methods, as is emphasized in [67]). Instead, the constraints are imposed on the intrinsic wavefunction, which can be readily handled using the method of Lagrange multiplier. Consequently, the intrinsic motion is still described by the original particle co-ordinates, and hence the intrinsic wavefunction is the shell-model wavefunction but subject to a well-defined constraint. Therefore, the intrinsic wavefunction permits us to use the available and well-known and extensive collection of shell-model concepts and tools.

Using a product wavefunction (i.e., a product of the shell-model wavefunction and the oscillation wavefunction) and an effective, separable, monopole-monopole interaction, and applying a constrained variational method to the transformed nuclear Schrödinger equation, we derive two separate but linked (via cranking and energy parameters) cranking-type Schrödinger equations, one for the oscillations and another for the intrinsic motion. A monopole constraint is imposed on the intrinsic wavefunction using a Lagrange multiplier and first order perturbation theory. The model parameters are dynamically and self-consistently determined by the solutions of the two equations themselves. For the harmonic oscillator mean-field potential, we solve the two equations exactly and determine the linking parameters and the excitation energy of the first excited $0^+$ state in the light nuclei as a function of the oscillation quantum number, intrinsic parameters, and the oscillation and intrinsic frequencies. A prescription using a sum rule is given for determining the oscillation frequency.

In Section 2, we derive the coupled oscillation-intrinsic Schrödinger equation from a transformation of the microscopic Bohr-Mottelson rotational model Schrödinger equation given in reference [60]. In Section 3, we apply to the Hamiltonian in Section 2 a constrained variational method to derive two separate cranking-type Schrödinger equations, one for the oscillations and another for the intrinsic motion, linked by the cranking parameters. In Sections 4, for the harmonic oscillator mean-field potential, we solve exactly the two equations derived in Section 3 to determine the eigenfunctions, energy eigenvalues, linking parameters, and excitation energy of the first excited $0^+$ states in the light nuclei. In Section 5, we present a sum rule prescription for determining the oscillation frequency. In Section 6, we apply the results of Sections 4 and 5 to predict the excitation energy of the first excited $0^+$ states in the light nuclei and compare the results to experimental data, with the goal of not reproducing the experimental data accurately but rather of predicting trends in the excitation energy and exploring the relative importance of the various features and approximations used in the model such as the constraint on the intrinsic wavefunction and the model parameters. Section 7 presents concluding remarks.

## 2. Derivation of coupled oscillation-intrinsic schrödinger equation

To derive the coupled oscillation-intrinsic Schrödinger equation in the three dimensional space, we start from the microscopic Bohr-Mottelson rotational-model [61] product wavefunction (as was done in [62] in the 2-*D* case):

$$\Psi_{JM} = \sum_{K=-J}^{J} \mathcal{D}_{MK}^{J}(\theta_s) \Phi_K(x_{ni}) \qquad (1)$$

where: $\mathcal{D}_{MK}^{J}$ is the Wigner rotation matrix, $J$, $\mathcal{M}$, and $K$ are respectively the total angular momentum (including spin) quantum number and its *z*-components along the space- and intrinsic-co-ordinate-system axes, $\theta_s$ ( $s =1,2,3$) are the three Euler angles specifying the orientation of the intrinsic axes, and $x_{nj}$ ($n = 1,..., A;\ j = 1,2,3,$ where $A$ = nuclear mass number) is a space-fixed nucleon co-ordinate. It is shown in [61] that the intrinsic wavefunction $\Phi_K$ in Eq. (1) satisfies the Schrödinger equation:

$$0 = \left( -\frac{\hbar^2}{2M} \sum_{n,j=1}^{A,3} \frac{\partial^2}{\partial x_{nj}^2} + \hat{V} - E \right) \Phi_{K'} + \frac{1}{2} \sum_{M,K=-J}^{+J} \sum_{A,B=1}^{3} \Phi_K \mathfrak{I}_{AB}^{rig^{-1}} \mathcal{D}_{MK'}^{J*} \hat{J}_B \hat{J}_A \mathcal{D}_{MK}^{J} \qquad (2)$$

(where $M$ is the nucleon mass, $\hat{J}_A$ is the $A^{th}$ component along the intrinsic axes of the total angular momentum operator, and $\mathfrak{I}_{AB}^{rig^{-1}}$ is the inverse of the rigid-flow moment of inertia tensor) provided the two-body interaction $\hat{V}$ and $\Phi_K$ are rotationally invariant, i.e.:

$$\left[ \hat{J}_j, \hat{V} \right] = 0, \qquad \hat{J}_A \Phi_K(x_{ni}) = 0 \qquad (3)$$

where $\hat{J}_j$ is the component of the total angular momentum along the $j^{th}$ space-fixed axis. Eq. (3) requires that $\Phi_K$ be a function of the rotation group *SO*(3) invariants such as $|\vec{x}_n|$, $|\vec{x}_n \cdot \vec{x}_m|$, $|\vec{x}_n - \vec{x}_m|$, etc., and/or be a zero angular momentum eigenstate.

For a description of oscillations (i.e., zero angular momentum state), we specialize Eq. (2) to a zero angular momentum state (i.e., set $J_j = 0$, $\mathcal{M} = 0, K = 0$) and obtain:

$$\left( -\frac{\hbar^2}{2M} \sum_{n,j=1}^{A,3} \frac{\partial^2}{\partial x_{nj}^2} + \hat{V} \right) \Phi = E \Phi \qquad (4)$$

(where for convenience we have dropped the zero subscript on $\Phi$).

As in [62], we now assume a product wavefunction of the form[1]:

$$\Phi = F(R) \cdot \phi(x_{ni}) \qquad (5)$$

where the oscillation co-ordinate (i.e., the nuclear radius) $R$ is defined by:

$$R \equiv \sum_{n=1}^{A} r_n^2 \equiv \sum_{n=1}^{A} (x_n^2 + y_n^2 + z_n^2) \qquad (6)$$

and $\phi$ is required to be independent of $R$, i.e.,

$$\frac{\partial}{\partial R} \phi = 0 \qquad (7)$$

We can take the intrinsic wavefunction $\phi$ (which is a function of the original independent particle co-ordinates $x_{nj}$) to be the shell model wavefunction but subject to the constraint in Eq. (7). The constraint can be interpreted to mean that the function $\phi$ is independent of $R$ and depends on the intrinsic co-ordinates only. An example of these intrinsic co-ordinates is the relative co-ordinates $x_{ni}'^2 \equiv x_{ni}^2 - R/(3A)$, analogously to the centre-of-mass case, with the properties $\sum_{n,k=1}^{A,3} x_{nk}'^2 = 0$ and $\partial x_{ni}'^2 / \partial R = 0$. In Section 5, we use the equality $\partial/\partial R = (3A)^{-1} \sum_{n,j=1}^{A,3} \partial/\partial x_{nj}^2 = (6A)^{-1} \sum_{n,j=1}^{A,3} (x_{nj})^{-1} \partial/\partial x_{nj}$ and a Lagrange multiplier to approximately implementation the constraint in Eq. (7).

---

[1] This ansatz is somewhat similar to that used in [58,59] for modelling giant monopole resonances, where the nuclear wavefunction is expanded in terms of an infinite series of hyperspherical or *K* harmonics with coefficients that are functions of the nuclear radius, and only the first term in the expansion is retained.

The transformation of Eq. (4) to the co-ordinate $R$ is readily achieved through the chain rule using Eqs. (5), (6), and (7) as follows:

$$\frac{\partial \Phi}{\partial x_{nj}} = \frac{\partial R}{\partial x_{nj}} \cdot \frac{dF}{dR} \phi + F \frac{\partial \phi}{\partial x_{nj}} = 2 x_{nj} \frac{dF}{dR} \phi + F \frac{\partial \phi}{\partial x_{nj}} \tag{8}$$

$$\sum_{n=1,j=1}^{A,3} \frac{\partial^2 \Phi}{\partial x_{nj}^2} = 2 \sum_{n=1,j=1}^{A,3} \frac{dF}{dR} \phi + 4 \sum_{n=1,j=1}^{A,3} x_{nj}^2 \frac{d^2 F}{dR^2} \phi + 4 \frac{dF}{dR} \cdot \sum_{n=1,j=1}^{A,3} x_{nj} \frac{\partial \phi}{\partial x_{nj}} + F \sum_{n=1,j=1}^{A,3} \frac{\partial^2 \phi}{\partial x_{nj}^2}$$

$$= \left( 6A \frac{dF}{dR} + 4R \frac{d^2 F}{dR^2} \right) \cdot \phi + 4 \frac{dF}{dR} \cdot \sum_{n=1,j=1}^{A,3} x_{nj} \frac{\partial \phi}{\partial x_{nj}} + F \sum_{n=1,j=1}^{A,3} \frac{\partial^2 \phi}{\partial x_{nj}^2} \tag{9}$$

Substituting Eq. (9) into Eq. (4) and using the identity:

$$\tilde{B} \equiv \frac{1}{2} \sum_{n,j=1}^{A,3} \left( x_{nj} \frac{\partial}{\partial x_{nj}} + \frac{\partial}{\partial x_{nj}} x_{nj} \right) = \frac{3A}{2} + \sum_{n,j=1}^{A,3} x_{nj} \frac{\partial}{\partial x_{nj}} \tag{10}$$

we obtain the coupled oscillation-intrinsic Schrödinger equation:

$$-4R \frac{d^2 F}{dR^2} \cdot \phi - 4 \frac{dF}{dR} \cdot \tilde{B} \phi + F \cdot \left( -\sum_{n=1}^{A} \nabla_n^2 + \frac{2M}{\hbar^2} \hat{V} \right) \cdot \phi = \frac{2ME}{\hbar^2} F \phi \equiv \varepsilon F \phi \tag{11}$$

where $\varepsilon$ is the reduced energy as defined on the right-hand side of Eq. (11). By making the transformation in Eqs. (5), (6), (7), and (9), we are seeking eigenvectors and eigenvalues of the nuclear Schrödinger Eq. (2) that correspond to coupled intrinsic-oscillation motion described by Eq. (11).

We now replace the interaction $\hat{V}$ by its Hartree-Fock mean field, which is commonly equated to the harmonic oscillator shell-model potential $\hat{V}_{os} = \frac{M\omega^2}{2} \sum_{m=1}^{A} r_m^2$. To obtain a potential (restoring-force) function for the monopole oscillations, we split $\hat{V}_{os}$ as follows:

$$\frac{2M}{\hbar^2} \hat{V}_{os} \equiv b^2 \cdot \sum_{m=1}^{A} r_m^2 \equiv b_m^2 \cdot R + b_s^2 \cdot \sum_{m=1}^{A} r_m^2 \tag{12}$$

where:

$$b \equiv \frac{M\omega}{\hbar}, \qquad \hbar\omega = 41 A^{-1/3} \tag{13}$$

In Eq. (13), we have used the empirically-determined expression for the oscillator frequency $\omega$ given in reference [68]. In this article we use the approximation $b_s = b$, and $b_m$ is determined in Section 5.

Replacing $\hat{V}$ in Eq. (11) by $\hat{V}_{os}$ in Eq. (12), we obtain the coupled oscillation-intrinsic Schrödinger equation:

$$\phi \cdot \left( -4R \frac{d^2}{dR^2} + b_m^2 R \right) \cdot F - 4 \frac{dF}{dR} \cdot \tilde{B} \phi + F \cdot \sum_{n=1}^{A} \left( -\nabla_n^2 + b_s^2 r_n^2 \right) \cdot \phi = \varepsilon F \phi \tag{14}$$

### 3. Derivation of self-consistent cranking model for oscillation-intrinsic motion

In this section, we apply the Rayleigh-Ritz variational method [51,52,69] to Eq. (14) subject to the normalization conditions:

$$\int \Phi^* \Phi \, d\tau = 1, \qquad \int_0^\infty F^* F \, d\tau_R = 1, \qquad \int_{-\infty}^\infty \phi^* \phi \, d\tau_{\text{int}} = 1 \tag{15}$$

(where $d\tau$'s are the body-fixed, oscillation and intrinsic volume elements, discussed below) to derive two coupled, self-consistent, cranking-type Schrödinger equations, one for the oscillation and another for the intrinsic motion. Accordingly, we take the expectation value of Eq. (14) using the product wavefuntion $\Phi = F \cdot \phi$ and minimize the (Lagrange multiplier) energy functional $\varepsilon = \varepsilon(F, \phi)$ with respect to arbitrary and separate variations $\delta F^*$ and $\delta \phi^*$. Note that it is not necessary to use both pairs of variations $\delta F^*$, $\delta \phi^*$, and $\delta F$, $\delta \phi$ because $\delta F^*$ and $\delta \phi^*$ are arbitrary [69].

Taking the expectation value of Eq. (14), we obtain:

$$\left(\langle F|-4R\frac{d^2}{dR^2}|F\rangle+b_m^2\langle F|R|F\rangle\right)\cdot\langle\phi|\phi\rangle-4\langle F|\frac{d}{dR}|F\rangle\cdot\langle\phi|\tilde{B}|\phi\rangle$$
$$+\left(\langle\phi|-\sum_{n=1}^{A}\nabla_n^2|\phi\rangle+b_s^2\langle\phi|\sum_{n=1}^{A}r_n^2|\phi\rangle\right)\cdot\langle F|F\rangle=\varepsilon\langle F|F\rangle\cdot\langle\phi|\phi\rangle \qquad(16)$$

Varying $F^*$ in Eq. (16), we obtain:

$$\langle\delta F|\left(-4R\frac{d^2}{dR^2}+b_m^2R-4\frac{d}{dR}\cdot\langle\phi|\tilde{B}|\phi\rangle+\langle\phi|-\sum_{n=1}^{A}\nabla_n^2|\phi\rangle+b_s^2\langle\phi|\sum_{n=1}^{A}r_n^2|\phi\rangle-\varepsilon\right)|F\rangle=0 \qquad(17)$$

where we have used the normalization in Eq. (15) and the energy minimization condition $\partial\varepsilon/\partial F^*=0$. Since the variation $\delta F^*$ is arbitrary, the quantity in the brackets in Eq. (17) must vanish, and we obtain the cranked monopole Schrödinger equation:

$$\left(-4R\frac{d^2}{dR^2}-4\langle\phi|\tilde{B}|\phi\rangle\cdot\frac{d}{dR}+b_m^2R\right)|F\rangle=\left(\varepsilon-t_{os}-b_s^2\langle\phi|\sum_{n=1}^{A}r_n^2|\phi\rangle\right)|F\rangle \qquad(18)$$

Similarly, varying $\phi^*$ in Eq. (16), we obtain the cranked intrinsic Schrödinger equation:

$$\left[-\sum_{n=1}^{A}\nabla_n^2-4\langle F|\frac{d}{dR}|F\rangle\cdot\tilde{B}+b_s^2\sum_{n=1}^{A}r_n^2\right]|\phi\rangle=\left(\varepsilon-t_R-b_m^2\langle F|R|F\rangle\right)|\phi\rangle \qquad(19)$$

In Eqs. (18) and (19) we have the definitions:

$$t_{os}\equiv\langle\phi|\left(-\sum_{n=1}^{A}\nabla_n^2\right)|\phi\rangle, \qquad t_R\equiv\langle F|\left(-4R\frac{d^2}{dR^2}\right)|F\rangle \qquad(20)$$

Taking the expectation value of either Eq. (18) or (19) using respectively $F$ and $\phi$, we obtain:

$$-4\langle\phi|\tilde{B}|\phi\rangle\cdot\langle F|\frac{d}{dR}|F\rangle=\varepsilon-t_{os}-t_R-b_m^2\langle F|R|F\rangle-b_s^2\langle\phi|\sum_{n=1}^{A}r_n^2|\phi\rangle \qquad(21)$$

From Eq. (21) we obtain the convenient definitions:

$$a\equiv-\frac{\beta_R}{4}\equiv\langle\phi|\tilde{B}|\phi\rangle=-\frac{1}{4}\cdot\frac{\varepsilon-t_{os}-t_R-b_m^2\langle F|R|F\rangle-b_s^2\langle\phi|\sum_{n=1}^{A}r_n^2|\phi\rangle}{\langle F|\frac{d}{dR}|F\rangle} \qquad(22)$$

$$\beta_{os}\equiv-4\langle F|\frac{d}{dR}|F\rangle=\frac{\varepsilon-t_{os}-t_R-b_m^2\langle F|R|F\rangle-b_s^2\langle\phi|\sum_{n=1}^{A}r_n^2|\phi\rangle}{\langle\phi|\tilde{B}|\phi\rangle} \qquad(23)$$

Substituting the definitions in Eqs. (22) and (23) into Eqs. (18), (19), and (21), we obtain:

$$\left(-4R\frac{d^2}{dR^2}+\beta_R\frac{d}{dR}+b_m^2R\right)|F\rangle=\left(\varepsilon-t_{os}-b_s^2\langle\phi|\sum_{n=1}^{A}r_n^2|\phi\rangle\right)|F\rangle \qquad(24)$$

$$\left(-\sum_{n=1}^{A}\nabla_n^2+\beta_{os}\cdot\tilde{B}+b_s^2\sum_{n=1}^{A}r_n^2\right)|\phi\rangle=\left(\varepsilon-t_R-b_m^2\langle F|R|F\rangle\right)|\phi\rangle \qquad(25)$$

$$a\cdot\beta_{os}=\varepsilon-t_{os}-t_R-b_m^2\langle F|R|F\rangle-b_s^2\langle\phi|\sum_{n=1}^{A}r_n^2|\phi\rangle \qquad(26)$$

Simultaneous solution of the three Eqs. (24), (25), and (26) together with the definitions in Eq. (20) and the normalization conditions in Eq. (15) determines the three unknown parameters $\varepsilon$, $a$, and $\beta_{os}$. (Note that Eqs. (22) and (23) are equivalent to any two of the Eqs. (24), (25) and (26), and hence Eqs. (22) and (23) are redundant and need not be considered further.)

Eqs. (24) and (25) may be viewed as a microscopic, self-consistent, cranking Schrödinger equations for the oscillation and intrinsic motions respectively, with the cranking parameters $\beta_R$ (or $a$) and $\beta_{os}$ being dynamical variables determined self-consistently by the two motions, as Eqs. (22) and (23) indicate. This self-consistency is further manifested by the obvious coupling between the two Eqs. (24) and (25). Another consequence of the self-consistency is that Eqs. (24) and (25) are time-reversal invariant because $a$ and $\beta_{os}$ are dynamical variables and must be chosen to have real values, as shown in Section 4. This is because the operators $d/dR$ and $\tilde{B}$ associated with $a$ and $\beta_{os}$ are real and non-hermitian unlike that in the conventional cranking model. These features require us to choose real (and not unitary) solutions of Eqs. (24) and (25). These features are improvements over the conventional phenomenological cranking models [2-6,9,24-55] where the cranking parameters are constant numbers and hence the models violate time-reversal invariance. It is clear that in Eq. (24) (and referring to Eq. (22)) the term $\beta_R d/dR$ represents the interaction between the oscillations and the averaged intrinsic single-particle motions. Similarly, in Eq. (25), the term $\beta_{os} \cdot \tilde{B}$ represents the interaction between the averaged oscillations and the intrinsic single-particle motions.

Eq. (24) is a microscopic, self-consistent, and arbitrary-amplitude quantum generalization of the phenomenological, semi-classical Bohr Hamiltonian for the vibrational motion of a spherical nucleus [1]. This connection becomes more transparent when $R$ is identified with the Bohr vibrational parameter $\beta^2$. As mentioned above, Eq. (24) also includes the interaction between the intrinsic and vibrational motions.

Another feature of Eqs. (24) and (25) is that each of the equations of motion has its own distinct cranking parameter, $a$ for the oscillatory motion and $\beta_{os}$ for the intrinsic motion. This differs from the conventional cranking model where the cranked Hamiltonian $H_{cr}$ is of the type: $H_{cr} = H_o - \dot{c}_{cr} \hat{O}_{cr}$, where $H_o$ is the intrinsic Hamiltonian, $\dot{c}_{cr}$ is the cranking parameter, and $\hat{O}_{cr}$ is the cranking operator (such as linear momentum, angular momentum, or dilation operator, refer to [2-6,9,24-55]). Eq. (25) may be considered to be a microscopic version of an equation that constrains the intrinsic system to have a definite mean value of $\tilde{B}$ (as in [55]). The above discussions and results provide the following microscopic prescription for rendering the conventional cranking model time-reversal invariant: replace $\hat{O}_{cr}$ by $\hat{O}_{cr}/(-i\hbar)$ in the conventional cranking model and use a real solution of the resulting equation. We note that besides rendering the cranked equations time-reversal invariant, the self-consistency also the determines the cranking parameters and the oscillation frequency $b_m$ as shown in Section 5.

Each of the Eqs. (24) and (25) determines completely the cranked wavefunction for that particular mode of cranked motion: Eq. (24) determines the cranked oscillation wavefunction and Eq. (25) the cranked intrinsic wavefunction, and the parameters in both equations are determined by the cranked wavefunctions (i.e. solutions) of both equations. The quantity on the right-hand-side of each of the Eqs. (24) and (25) is the energy eigenvalue of that mode of cranked motion. The parameter $\varepsilon$ is the total energy of the combined system.

## 4. Solution of self-consistent cranking oscillation-intrinsic equations

We want to solve Eqs. (24), (25), and (26), together with the definitions in Eq. (20) and the normalization conditions in Eq. (15), for the three unknowns $\varepsilon$, $a$, and $\beta_{os}$ and express them in terms of $b_m$, $b_s = b$, intrinsic-system parameters, and oscillation excitation quantum number.

*4.1  Solution of cranked intrinsic schrödinger Eq. (25)*

We now obtain the eigenvalues and eigenfunctions of Eq. (25). We observe that the operator $\tilde{B}$ in Eqs. (25) and (10) is not self-adjoint or hermitian. This means that the differential Eq. (25) is not self-adjoint and hence its solution is not unitarily related to the harmonic oscillator eigenfunctions. Of-course, we can make Eq. (25) self-adjoint by multiplying both sides of Eq. (25) by $\hbar^2/2M$ and replacing $\tilde{B}$ by $\hat{B} \equiv -i\hbar\tilde{B}$, and $\beta_{os}$ by $\bar{\beta} \equiv i\hbar\beta_{os}$, and obtain (from a direct solution of the differential equation or from an algebraic approach using its underlying dynamical non-compact Lie algebra $su(1,1)$) a unitary solution of the resulting Eq. (25) as in the conventional cranking or squashing model [55]. This unitary solution yields a real value for the expectation of the operator $\hat{B}$ since $\hat{B}$ is hermitian, and an imaginary value for the expectation of the operator $\tilde{B}$ in Eq. (10) and for $\beta_{os}$ in Eqs. (23) and (25) and for $\beta_R$ in Eqs. (22) and (24). Imaginary $\beta_{os}$ and $\beta_R$ results in unphysical solutions to Eqs. (24) and (25). Therefore, a unitary solution of Eq. (25) is unacceptable for the solution of the coupled Eqs. (24) and (25). One can show that the energy eigenvalues for unitary and real solutions of Eq. (25) are the same.

These results underscore the importance of self-consistency in the cranking model. Self-consistency rejects unitary solutions of the cranked intrinsic Eq. (25) and requires a solution of Eq. (25) that yields a real value for the cranking parameter $\beta_R$, and this is intimately related to time-reversal invariance of the coupled and self-consistent intrinsic-oscillation Eqs. (24) and (25). In a conventional cranking model, which violates time-reversal invariance, such as any form of the Inglis's model [2-6,9,24-55], there is no reason to reject a unitary solution of the equation of motion because the cranking parameter is a constant and is not determined self-consistently by the coupling between the intrinsic and collective motions.

We, therefore, want a real solution of Eq. (25) to yield a real value of $\beta_R$. This real solution is derived and is given by the Slater determinant:

$$\phi = \frac{1}{\sqrt{A!}} \hat{P} \prod_{n=1}^{A} \varphi_n(x_n) \cdot \varphi_n(y_n) \cdot \varphi_n(z_n) \tag{27}$$

$$\varphi_n(x) = c_n e^{-\beta_c \tilde{x}^2/2} H_n(\tilde{x}), \qquad n = 0,1,2,\ldots\infty \tag{28}$$

where $\hat{P}$ is the particle-occupation permutation or anti-symmetrization operator, and:

$$c_n \equiv \left[\frac{b_c}{2^{2n}\pi(n!)^2}\right]^{-1/4}, \quad b_c \equiv \sqrt{b_s^2 + \beta_{os}^2/4}, \quad \beta_c \equiv 1 - \frac{\beta_{os}}{2b_c}, \quad \tilde{x} \equiv \sqrt{b_c} \cdot x \tag{29}$$

$c_n$ is the normalization constant, and $H_n$ is the Hermite polynomial. The eigenfunctions $\varphi_n$ are orthonormal with respect to the weighting function or measure $w_{cos} \equiv e^{-\beta_{os} x^2/2}$, which is derived from an application of Sturm-Liouville boundary value problem [70] to Eq. (25):

$$\int_{-\infty}^{\infty} w_{cosx} \varphi_m \cdot \varphi_n \, dx = \int_{-\infty}^{\infty} e^{-\beta_{os} x^2/2} \varphi_m \cdot \varphi_n \, dx = \delta_{nm}, \quad \text{and hence} \quad \int_{-\infty}^{\infty} e^{-\beta_{os} \sum_{m=1}^{A} r_m^2/2} \phi^2 \cdot \prod_{n,k=1}^{A,3} dx_{nk} = 1 \tag{30}$$

for any positive or negative value of $\beta_{os}$. The measure $w_{cos} \equiv e^{-\beta_{os}\sum_{m=1}^{A} r_m^2/2}$ in Eq. (30) also renders Eq. (25) self-adjoint (hermitian).

The eigenvalues of Eq. (25) are determined to be:

$$\varepsilon - t_R - b_m^2 \langle F|R|F\rangle = 2\Sigma\sqrt{b_s^2 + \beta_{os}^2/4} \tag{31}$$

where the total particle-occupation number $\Sigma$ is:

$$\Sigma \equiv \sum_{n,k=1}^{n_f,3}\left(n_k + \frac{1}{2}\right) \tag{32}$$

and $n_f$ is the Fermi level oscillator quantum number.

The quantity $t_{os}$ is evaluated with respect to the measure $w_{cos}$ in Eq. (20) to be given by:

$$t_{os} = \int_{-\infty}^{\infty} e^{-\beta_{os}\sum_{m=1}^{A} r_m^2/2} \cdot \phi\left(-\sum_{n=1}^{A}\nabla_n^2\right)\phi \cdot \prod_{n,k=1}^{A,3} dx_{nk} = \frac{b_s^2 \cdot \Sigma}{\sqrt{b_s^2 + \beta_{os}^2/4}} \tag{33}$$

Similarly, $\langle\phi|\sum_{n=1}^{A} r_n^2|\phi\rangle$ in Eqs. (21), (24), and (26) is evaluated to be:

$$\langle\phi|\sum_{n=1}^{A} r_n^2|\phi\rangle = \frac{\Sigma}{\sqrt{b_s^2 + \beta_{os}^2/4}} \tag{34}$$

Note that each of the oscillator eigenfunctions in Eq. (28) is a superposition of the original (uncranked) oscillator eigenfunctions, and hence the cranked oscillator eigenvalues correspond to a re-ordering of (i.e., are different from) the uncranked oscillator eigenvalues even though the particle occupation number $\Sigma$ in Eq. (32) has the same value for both cases. This is because the original and cranked oscillator frequencies differ from each other ($b_s$ versus $b_c$, refer to Eq. (29)).

## 4.2  Solution of cranked oscillation schrödinger Eq. (24)

In this section, we obtain the eigenfunctions and eigenvalues of Eq. (24). Eq. (24) resembles Eq. (18) in [62], and therefore we can determine its unitary eigenfunctions and eigenvalues using its underlying dynamical non-compact Lie algebra $su(1,1)$. However, the analysis for this determination is a local rather than global analysis, and hence it does not allow us to evaluate some of the integrals occurring in the parameters in Eqs. (20), (24), (25) and (26). Furthermore, these unitary solutions do not yield real values for the parameter $t_R$ in Eq. (20).

For these reasons, we determine explicitly (i.e., globally) from the literature [69,70,71,72] the real eigenfunctions and eigenvalues of differential Eq. (24). One set of eigenfunctions of Eq. (24) is given by (the other set does not yield physical results and hence is not considered):

$$F_n = d_n e^{-\tilde{R}/2} K(-n, a, \tilde{R}), \quad n = 0, 1, 2, 3, \ldots \infty, \quad \tilde{R} \equiv b_m \cdot R \tag{35}$$

where $d_n$ is a normalization factor, $a$ is defined in Eq. (22), and $K$ is the (polynomial or finite series) confluent hypergeometric or Kummer function [69,70,71,72]. The eigenfunctions $F_n$ are orthonormal with respect to the weighting function or measure $w \equiv R^{a-1}$. The parameter $a$ must satisfy the condition:

$$a > 1, \quad \text{or} \quad \beta_R < -4 \tag{36}$$

to ensure that $F_n$ are orthogonal and certain integrated quantities vanish on the boundaries. The eigenvalues corresponding to the eigenfunctions in Eq. (35) are:

$$\varepsilon - t_{os} - b_s^2 \langle \phi | \sum_{n=1}^{A} r_n^2 | \phi \rangle = 2 b_m \cdot (2n + a) \tag{37}$$

Note that the energy eigenvalues in Eq. (37) are identical to those obtained from the unitary $su(1,1)$ solution of Eq. (24).

Using the measure $w \equiv R^{a-1}$ and the function in Eq. (35) (more precisely its generating function), we obtain the following expression for the parameters $t_R$ and $\langle F|R|F \rangle$ in Eqs. (20), (24), (25), and (26):

$$t_R = \int_0^\infty R^{a-1} F_n \left(-4 \frac{d^2}{dR^2}\right) F_n \, dR = b_m \cdot (2n + a - \mathcal{B}_n) \tag{38}$$

$$\langle F|R|F \rangle = \int_0^\infty R^{a-1} F_n R F_n \, dR = \frac{2n + a}{b_m} \tag{39}$$

where:

$$\mathcal{B}_n \equiv \frac{2 \cdot n!}{(a+1)(a+2)\ldots(a+n-1)} \sum_{k=0}^{n} \frac{(a-1)a\ldots(a+n-k-2)}{(n-k)!} \tag{40}$$

Note that $\mathcal{B}_n = 2$ for $n = 0$ and $\mathcal{B}_n = 2a$ for $n = 1$.

## 4.3  Self-consistent determination of parameters $\varepsilon$, $a$, and $\beta_{os}$

We now combine Eqs. (26), (31), (33), (34), (37), (38), (39), and (40) to express $\varepsilon$, $a$, and $\beta_{os}$ in terms of $b_m$, $b_s$, $\Sigma$, and oscillation excitation quantum number $n$. Subtracting Eq. (38) from Eq. (37) and using Eq. (34), we obtain:

$$\varepsilon - t_{os} - t_R = b_m \cdot (2n + a + B_n) + \frac{b_s^2 \cdot \Sigma}{\sqrt{b_s^2 + \beta_{os}^2/4}} \tag{41}$$

Substituting Eq. (41) into Eq. (26) and using Eqs. (34) and (39), we obtain:

$$a \cdot \beta_{os} = b_m \cdot (2n + a + \mathcal{B}_n) + \frac{b_s^2 \cdot \Sigma}{\sqrt{2b_s^2 + \beta_{os}^2/4}} - b_m \cdot (2n + a) - \frac{b_s^2 \cdot \Sigma}{\sqrt{2b_s^2 + \beta_{os}^2/4}} = b_m \cdot \mathcal{B}_n \qquad (42)$$

From Eq. (42) (noting that $\mathcal{B}_n = 2$ for $n = 0$ and $\mathcal{B}_n = 2a$ for $n = 1$), we obtain:

$$b_m = \frac{a \cdot \beta_{os}}{2} \quad \text{for } n = 0, \quad \text{and} \quad \beta_{os} = 2b_m \quad \text{for } n = 1 \qquad (43)$$

Subtracting Eq. (33) from Eq. (31), using Eq. (39), and equating the resulting equation to Eq. (41), we obtain:

$$\varepsilon - t_{os} - t_R = 2\Sigma\sqrt{b_s^2 + \beta_{os}^2/4} + b_m \cdot (2n + a) - \frac{b_s^2 \cdot \Sigma}{\sqrt{b_s^2 + \beta_{os}^2/4}} = b_m \cdot (2n + a + \mathcal{B}_n) + \frac{b_s^2 \cdot \Sigma}{\sqrt{b_s^2 + \beta_{os}^2/4}} \qquad (44)$$

From Eq. (44) we obtain:

$$b_m \cdot \mathcal{B}_n = \frac{\beta_{os}^2 \cdot \Sigma}{2\sqrt{b_s^2 + \beta_{os}^2/4}} \qquad (45)$$

For $n = 0$ (noting that $\mathcal{B}_n = 2$ for n = 0), Eq. (45) gives:

$$b_m = \frac{\beta_{os}^2 \cdot \Sigma}{2\sqrt{b_s^2 + \beta_{os}^2/4}} \qquad (46)$$

Solving Eq. (46) for $\beta_{os}$, we obtain:

$$\beta_{os}^2 = \frac{2b_m^2}{\Sigma^2}\left(1 + \sqrt{1 + \frac{4b_s^2 \Sigma^2}{b_m^2}}\right) \qquad (47)$$

where we have chosen the positive sign to ensure positive $\beta_{os}^2$. Substituting Eq. (47) into Eq. (43) for $n = 0$, we obtain:

$$a = 2\Sigma\left(1 + \sqrt{1 + \frac{4b_s^2 \Sigma^2}{b_m^2}}\right)^{-1/2} \qquad (48)$$

For $n = 1$, noting that $\mathcal{B}_n = 2a$ and using Eq. (43) for $n = 1$, Eq. (45) gives:

$$a = \frac{\beta_{os}^2 \cdot \Sigma}{4b_m\sqrt{b_s^2 + \beta_{os}^2/4}} = \frac{b_m \cdot \Sigma}{\sqrt{b_s^2 + b_m^2}} \qquad (49)$$

The total reduced energy $\varepsilon$ is obtained from Eqs. (31), (38), and (39) (or from Eqs. (33), (34), and (37)):

$$\varepsilon = 2\Sigma\sqrt{b_s^2 + \beta_{os}^2/4} + 2b_m \cdot \left(2n + a - \frac{\mathcal{B}_n}{2}\right) \qquad (50)$$

The excitation energy for the first excited $0^+$ state is naturally defined as follows (recalling the definition of reduced energy in Eq. (11)):

$$\Delta E_m \equiv \frac{\hbar^2}{2M}\left[\varepsilon(n=1) - \varepsilon(n=0)\right] = \frac{\hbar^2 b}{2M} \cdot \frac{\varepsilon(n=1) - \varepsilon(n=0)}{b} = \frac{\hbar\omega}{2} \cdot \frac{\varepsilon(n=1) - \varepsilon(n=0)}{b} \qquad (51)$$

where we have used the definition of $b$ in Eq. (13) and have non-dimensionalized the reduced energy $\varepsilon$ by dividing it by $b$, and hence readily expressible in terms of the following non-dimensional quantities:

$$\tilde{b}_m \equiv \frac{b_m}{b}, \quad \tilde{b}_s \equiv \frac{b_s}{b}, \quad \tilde{\beta}_{os} \equiv \frac{\beta_{os}}{b} \qquad (52)$$

## 5. Sum-rule prescription for determining oscillation frequency $b_m$

In this section, we use an energy-weighted sum-rule and the intrinsic-system constraint in Eq. (7) to compute the oscillation frequency $b_m$ and hence the oscillation excitation energy in Eq. (51)[2].

---

[2] One may relate the oscillation excitation energy to the nuclear compressibility, ground-state energy, and the

First we note that Eq. (48) and the condition in Eq. (36) place the following limit on the range of values of $\tilde{b}_m$ (noting that $\Sigma \geq 6$ for the light nuclei listed in Table 1):

$$a = 2\Sigma \left(1 + \sqrt{1 + \frac{4b_s^2 \Sigma^2}{b_m^2}}\right)^{-1/2} > 1 \quad \Rightarrow \quad \tilde{b}_m > \frac{2\Sigma}{\sqrt{(2\Sigma^2 - 1)^2 - 1}} > \frac{1}{\Sigma} \leq 0.17 \tag{53}$$

The energy-weighted sum rule [7,9-16] is given by the identity relationship (for any operator $\hat{A}$ and a complete set of states $|v\rangle$ with energies $E_v$ corresponding to a Hamiltonian $H$):

$$\sum_{v=0}(E_v - E_o)|\langle v|\hat{A}|0\rangle|^2 = \sum_{v=0}(E_v - E_o)\langle 0|\hat{A}|v\rangle\langle v|\hat{A}|0\rangle = \frac{1}{2}\langle 0|[[\hat{A},H],\hat{A}]|0\rangle \tag{54}$$

where we have used the completeness property $\sum_{v=0}|v\rangle\langle v| = 1$. For the Hamiltonian $H_s$ given in Eq. (25), (which is state dependent, i.e. a functional) with energy $E \equiv \hbar^2\varepsilon/2M$, namely:

$$H_s \equiv \frac{\hbar^2}{2M}\left(-\sum_{n=1}^{A}\nabla_n^2 + \beta_{os}\cdot\tilde{B} + b_s^2\sum_{n=1}^{A}r_n^2 + t_{Rn} + b_m^2\langle F_n|R|F_n\rangle\right)$$

where $t_{Rn} \equiv \langle F_n| -4R\frac{d^2}{dR^2}|F_n\rangle$ and $|v\rangle = |F_n\rangle$, Eq. (54) becomes:

$$\sum_{n=0}(E_n - E_o)|\langle\phi|\langle F_n|\hat{A}|F_o\rangle|\phi\rangle|^2 = \frac{1}{2}\langle\phi|\langle F_o|\left[\left[\hat{A},\frac{\hbar^2}{2M}\left(-\sum_{n=1}^{A}\nabla_n^2 + \beta_{os}\cdot\tilde{B}\right)\right],\hat{A}\right]|F_o\rangle|\phi\rangle$$
$$= \frac{\hbar^2}{2M}\langle\phi|\langle F_o|\sum_n \vec{\nabla}_n\hat{A}\cdot\vec{\nabla}_n\hat{A}|F_o\rangle|\phi\rangle \tag{55}$$

We choose $\hat{A} = \sum_{n=1}^{A} r_n^2 = R$, and Eq. (55) gives:

$$\sum_{n=0}(E_n - E_o)|\langle\phi|\langle F_n|R|F_o\rangle|\phi\rangle|^2 = \frac{2\hbar^2}{M}\langle\phi|\sum_{n=1}^{A}r_n^2|\phi\rangle \tag{56}$$

For the Hamiltonian $H_m$ given in Eq. (24) with energy $E \equiv \hbar^2\varepsilon/2M$, namely:

$$H_m \equiv \frac{\hbar^2}{2M}\left(-4R\frac{d^2}{dR^2} + \beta_R\frac{d}{dR} + b_m^2 R + t_{os} + b_s^2\langle\phi|\sum_{n=1}^{A}r_n^2|\phi\rangle\right) \tag{57}$$

Eq. (55) becomes:

$$\sum_{n=0}(E_n - E_o)|\langle\phi|\langle F_n|R|F_o\rangle|\phi\rangle|^2 = \frac{\hbar^2}{4M}\langle F_o|\left[\left[R, -4R\frac{d^2}{dR^2} + \beta_R\frac{d}{dR}\right], R\right]|F_o\rangle = \frac{2\hbar^2}{M}\langle F_o|R|F_o\rangle \tag{58}$$

Equating the right-hand sides of Eqs. (56) and (58), we obtain:

$$\langle\phi|\sum_{n=1}^{A}r_n^2|\phi\rangle = \langle F_o|R|F_o\rangle \tag{59}$$

Substituting Eqs. (34) and (39) into Eq. (59), we obtain:

$$\frac{\Sigma}{\sqrt{b_s^2 + \beta_{os}^2(n=0)/4}} = \frac{a(n=0)}{b_m} \tag{60}$$

Squaring both sides of Eq. (60) and substituting Eqs. (47) and (48) into the resulting equation, we obtain (for $\Sigma \geq 6$):

$$\tilde{b}_m \simeq \frac{\Sigma}{\Sigma^2 - 1} > \frac{1}{\Sigma} \tag{61}$$

---

nuclear radius $R$ using small-amplitude oscillation concepts [7,8,11,15,20,21,22,73-76] and thereby derive an expression for the oscillation frequency. We are not using this approach in this article because it predicts unrealistically low value for $\tilde{b}_m$.

The result in Eq. (61) places a slightly higher value on the lower bound of $\tilde{b}_m$ than that given in Eq. (53).

To obtain a more realistic value for $\tilde{b}_m$, we have approximately imposed the constraint in Eq. (7) on the intrinsic wavefunction $|\phi\rangle$ in the form of the expectation (i.e., the first moment) of the constraint in Eq. (7) and included it in the formulation in Sections 3 to 4.2 using the method of Lagrange multiplier and first-order perturbation theory. For $^4_2He$, we have cast the results of the perturbation analysis into the following convenient form[3] for the quantities in Eqs. (31), (33), and (34):

$$\varepsilon - t_R - b_m^2 \langle F|R|F\rangle = 2\Sigma^{1.08}\sqrt{b_s^2 + \beta_{os}^2/4}, \quad t_{os} = \frac{b_s^2 \cdot \Sigma^{1.17}}{\sqrt{b_s^2 + \beta_{os}^2/4}}, \quad \langle\phi|\sum_{n=1}^{A} r_n^2|\phi\rangle = \frac{\Sigma^{0.8}}{\sqrt{b_s^2 + \beta_{os}^2/4}} \tag{62}$$

Using Eqs. (62) and steps given in Section 4.2, we obtain:

$$\beta_{os}^2(n=0) = \frac{2b_m^2}{\overline{\Sigma}^2}\left(1 + \sqrt{1 + \frac{4b_s^2 \overline{\Sigma}^2}{b_m^2}}\right) \qquad a(n=0) = \frac{2b_m}{\beta_{os}} = \sqrt{2}\overline{\Sigma}\left(1 + \sqrt{1 + \frac{4b_s^2 \overline{\Sigma}^2}{b_m^2}}\right)^{-1/2}$$

$$\beta_{os}(n=1) = 2b_m \qquad a(n=1) = \frac{b_m \overline{\Sigma}}{\sqrt{b_s^2 + b_m^2}} \tag{63}$$

where $\overline{\Sigma} \equiv \Sigma^{1.08}$. Squaring both side of Eq. (60) and substituting Eq. (63) into the resulting equation, we obtain:

$$\tilde{b}_m \simeq \Sigma^{-0.52} \tag{64}$$

We also use the approximate result in Eq. (64) for the other light nuclei because the predictions in this article are of exploratory nature and more rigorous study will be conducted in a future article.

## 6. Preliminary assessment of model for light nuclei

In this section, we perform a preliminary scoping calculation of the excitation energy of the first excited $0^+$ state in the light nuclei in the range $A \leq 120$ using the mean-field harmonic oscillator potential and the equations derived in Sections 4.2 and 5. However, the main objective and the result of the analysis in this article are the derivation of the microscopic, quantal, self-consistent cranking model and its self-consistently determined cranking parameters, which are presented in Sections 3, 4, and 5.

Since presently the model uses first moment of the intrinsic constraint, neglects the centre-of-mass motion, the residual two-body interaction (in such forms as pairing and Nilsson's shell-model corrections), and deformation, this preliminary calculation is not intended to accurately predict this excitation energy but rather to show the predicted trends and explore the relative importance of the various features, the approximations (such as the constraint on the intrinsic system and its first-order treatment), and the parameters used in the model. In particular, the residual interaction in any of its (microscopic or phenomenological) forms can have a significant impact on the excitation energy as indicated in [58]. Therefore, without the inclusion of a residual interaction, we do not expect a reasonable agreement with the experimental data although we may have a better success in predicting the excitation energy of giant monopole resonances (as shown in [58]) and quadrupole vibration excitation (when nuclear deformation is included) since these excitation energies may be less sensitive to the neglected factors.

We have used the Microsoft Office Excel program and Eqs. (13), (51), (62), (63), and (64) to compute, for $a > 1$, the excitation energy of the first excited $0^+$ state in several light nuclei in the range

---

[3] In deriving these results, we have dropped terms in the perturbation-theory formula for the Lagrange multiplier corresponding to intrinsic-system particle-hole excitations higher than $4\hbar\omega$.

$A \leq 120$ as follows using the dimensionless $\tilde{b}_m$ in Eq. (52). For each of these nuclei, i.e., for given values of $\omega$ (in Eq. (13)), and $\Sigma$ (in Eq. (32)), we choose the value of $\tilde{b}_m$ computed from Eq. (64), and compute $\beta_{os}$ and $a$ in Eq. (63), $\varepsilon$ in Eq. (62), and $\Delta E_m$ in Eq. (51). We assume in this article that the cranked intrinsic system remains in its ground state, i.e., we use the lowest value of $\Sigma$ for each of the nuclei (given in Table 1). We have set $b_s = b$ and used the empirical prescription in Eq. (13) for the oscillator frequency $\omega$. We have neglected: the centre-of-mass motion and residual two-body interaction, such particle-pairing interaction, which may be important in the oscillations [58,77], a realistic shell model such as Nilsson's model [68], and the constraint in Eq. (3), which is mostly taken care of since we are dealing only with even-even nuclei where each single-particle orbit is filled with a pair of particles with spins pointing in the opposite direction.

**Table 1.** Isotropic-oscillator ground- (lowest-energy) state particle-occupation configurations[4].

$^{4}_{2}He: (000)^4$, $\Sigma = 6$

$^{12}_{6}C: (000)^4 (010)^4 (001)^4$, $\Sigma = 26$

$^{16}_{8}O: (000)^4 (100)^4 (010)^4 (001)^4$, $\Sigma = 36$

$^{28}_{14}Si: {}^{16}_{8}O\ (110)^4 (101)^4 (011)^4$, $\Sigma = 78$

$^{32}_{16}S: {}^{28}_{14}Si\ (200)^4$, $\Sigma = 92$

$^{40}_{20}Ca: {}^{32}_{16}S\ (020)^4 (002)^4$, $\Sigma = 120$

$^{60}_{28}Ni: {}^{40}_{20}Ca\ (300)^4 (111)^4 (210)^4 (201)^4 (102)^2 (120)^2$, $\Sigma = 210$

$^{62}_{28}Ni: {}^{60}_{28}Ni\ (012)^2$, $\Sigma = 219$

$^{70}_{32}Ge: {}^{40}_{20}Ca\ (300)^4 (111)^4 (210)^4 (201)^4 (102)^4 (120)^4 (012)^2 (021)^2 (030)^2$, $\Sigma = 255$

$^{90}_{40}Zr: {}^{40}_{20}Ca\ (300)^4 (111)^4 (210)^4 (201)^4 (102)^4 (120)^4 (012)^4 (021)^4 (030)^4 (003)^4\ (400)^2 (211)^2 (112)^2 (121)^2 (022)^2$, $\Sigma = 355$

$^{114}_{48}Cd: {}^{40}_{20}Ca\ (300)^4 (111)^4 (210)^4 (201)^4 (102)^4 (120)^4 (012)^4 (021)^4 (030)^4 (003)^4 (400)^4 (211)^4 (112)^4$ $(121)^4 (022)^2 (202)^2 (220)^2 (310)^2 (301)^2 (130)^2 (031)^2 (103)^2 (411)^2$, $\Sigma = 491$

$^{116}_{50}Sn: {}^{40}_{20}Ca\ (300)^4 (111)^4 (210)^4 (201)^4 (102)^4 (120)^4 (012)^4 (021)^4 (030)^4 (003)^4\ (400)^4 (211)^4 (112)^4$ $(121)^4 (022)^4 (202)^2 (220)^2 (310)^2 (301)^2 (130)^2 (031)^2 (103)^2 (411)^2$, $\Sigma = 502$

$^{120}_{52}Te: {}^{40}_{20}Ca\ (300)^4 (111)^4 (210)^4 (201)^4 (102)^4 (120)^4 (012)^4 (021)^4 (030)^4 (003)^4 (400)^4 (211)^4 (112)^4$ $(121)^4 (022)^4 (202)^4 (220)^2 (310)^2 (301)^2 (130)^2 (031)^2 (103)^2 (411)^2 (040)^2$, $\Sigma = 524$

Table 2 presents the model-predicted ($\Delta E_m$) and the experimentally-observed ($\Delta E_{exp}$) excitation energies and the other model-predicted parameters in the light nuclei for $a > 1$.

Table 2 shows that the present model predicts reasonably well the excitation energy $\Delta E_m$ of the first excited $0^+$ state in $^{4}_{2}He$ and $^{12}_{6}C$. The accuracy of the prediction is surprising and may be fortuitous in view of the results in [58] (that emphasize the importance of the two-body interaction in monopole oscillations) and will be examined in a future article where the neglected factors mentioned above will be included in the model. Nevertheless, this result seem to indicate that this state is mostly spherical and of monopole nature [78,79,80] since these two nuclei are observed to have no neighbouring rotational or quadrupole vibrational states [81] and hence have spherical shape in the first excited $0^+$ state. In $^{16}_{8}O$, $\Delta E_m$ is overpredicted by 13%, which is expected because it is known experimentally and analytically that the first excited $0^+$ state in $^{16}_{8}O$ forms the start (or the bandhead) of the ground-state rotational band [18,57,68,82-86], and hence the state is deformed and undergoes quadrupole beta-vibrational and rotational motion. Deformation and the resulting vibrational and rotational motion lowers the excitation energy below that of the monopole mode. $\Delta E_m$ is reasonably well predicted in $^{28}_{14}Si$. This result is again surprising (and will be further studied in a future article) because the first excited $0^+$ state in $^{28}_{14}Si$ is known experimentally and analytically to form the bandhead of the ground-state rotational band [57,80,87], and hence this state may be slightly deformed. The model overpredicts by 21% the $\Delta E_m$ in $^{32}_{16}S$. This may not be surprising since experimentally the first excited $0^+$ state in $^{32}_{16}S$ seems to form a member of a quadrupole (beta) vibrational triplet [81]. The model also overpredicts by 21% the $\Delta E_m$ in $^{40}_{20}Ca$ since the first excited $0^+$ state in this nucleus seems to form the bandhead of the ground-state rotational band [81,87,88].

---

[4] $(n_x n_y n_z)^m$ indicates oscillator-state quantum numbers in the $x$, $y$, and $z$ directions, and the superscript $m$ indicates the number of nucleons in this orbit, and $^{28}_{14}Si\ ^{28}_{14}Si\ (200)^4$, for example, indicates the combined configuration

    The model overpredicts the $\Delta E_m$ in $^{60}_{28}Ni$ and $^{62}_{28}Ni$ by 48% and 57% respectively since experimentally the first excited $0^+$ state in each of $^{60}_{28}Ni$ and $^{62}_{28}Ni$ seems to form a member of a quadrupole (beta) vibrational triplet [81]. The model overpredicts by 56% the $\Delta E_m$ in $^{90}_{40}Zr$ since the first excited $0^+$ state in this nucleus seems to form the bandhead of the ground-state rotational band [81,88]. The model overpredicts the $\Delta E_m$ in $^{70}_{32}Ge$, $^{114}_{48}Cd$, and $^{120}_{52}Te$ by 167%, 136%, and 127% respectively because the first excited $0^+$ state in each of these nuclei seems to form a member of a quadrupole (beta) vibrational triplet [81,87,88]. In $^{116}_{50}Sn$ $\Delta E_m$ is overpredicted by 44% perhaps because its first excited $0^+$ state may be slightly quadrupole-beta deformed and hence undergoes quadrupole beta-vibrational and rotational motion as evidenced by the electric quadrupole ($E$2) transition to neighbouring $2^+$ state.

**Table 2.** Predicted ($\Delta E_m$) and Observed ($\Delta E_{exp}$) first excited $0^+$ state excitation energy in light nuclei.

| Nucleus | Ang. mom./parity | Multi-polarity | $\tilde{b}_m$ | $a$ $n=0/n=1$ | $\tilde{\beta}_{os}$ $n=0/n=1$ | $\Delta E_m$ (MeV) [5] (Eqs. 55, 88) | $\Delta E_{exp}$ (MeV) |
|---|---|---|---|---|---|---|---|
| $^{4}_{2}He$ | $0^+$ | $E0$ | 0.3656 | 1.657/2.38 | 0.47/0.73 | 20.3 (1%) | 20.1 |
| $^{12}_{6}C$ | $0^+$ | $E0$ | 0.1605 | 2.32/5.35 | 0.14/0.32 | 8.2 (6%) | 7.7 |
| $^{16}_{8}O$ | $0^+$ | $E0$ | 0.1337 | 2.53/6.35 | 0.11/0.27 | 6.9 (13%) | 6.1 |
| $^{28}_{14}Si$ | $0^+$ | $E0$ | 0.0866 | 3.09/9.53 | 0.06/0.17 | 4.9 (-1%) | 4.97 |
| $^{32}_{16}S$ | $0^+$ | $E0$ | 0.0789 | 3.23/10.39 | 0.05/0.16 | 4.6 (21%) | 3.8 |
| $^{40}_{20}Ca$ | $0^+$ | $E0$ | 0.068 | 3.46/11.94 | 0.04/0.14 | 4.1 (21%) | 3.4 |
| $^{60}_{28}Ni$ | $0^+$ | $E0$ | 0.0496 | 4.0/15.97 | 0.03/0.10 | 3.4 (48%) | 2.3 |
| $^{62}_{28}Ni$ | $0^+$ | $E0$ | 0.0485 | 4.04/16.32 | 0.02/0.10 | 3.3 (57%) | 2.1 |
| $^{70}_{32}Ge$ | $0^+$ | $E0$ | 0.0445 | 4.21/17.67 | 0.02/0.09 | 3.2 (167%) | 1.2 |
| $^{90}_{40}Zr$ | $0^+$ | $E0$ | 0.037 | 4.59/20.98 | 0.02/0.07 | 2.8 (56%) | 1.8 |
| $^{114}_{48}Cd$ | $0^+$ | $E0/E2$ | 0.0308 | 4.98/24.82 | 0.01/0.06 | 2.6 (136%) | 1.1 |
| $^{116}_{50}Sn$ | $0^+$ | $E0/E2$ | 0.0304 | 5.01/25.11 | 0.01/0.06 | 2.6 (44%) | 1.8 |
| $^{120}_{52}Te$ | $0^+$ | decay to final states $0^+/2^+$ | 0.0297 | 5.07/25.68 | 0.01/0.06 | 2.5 (127%) | 1.1 |

## 7. Concluding remarks

The eventual objective of the analysis in this article is to develop a microscopi, quantal, self-consistent cranking model for vibrational-rotational motion in deformed nuclei. To this end, it is instructional to and we develop, in this article, the simpler cranking model for oscillations in a spherical nucleus.

Accordingly, we transform and thereby decompose the Bohr-Mottelson rotational model Hamiltonian developed previously into the sum of three parts, one describing nuclear radius oscillations, another describing the intrinsic motion, and the third part describing the coupling between these two motions. In this approach we do not use redundant co-ordinates and/or place any constraints on the particle co-ordinates. Instead, the constraints are imposed on the intrinsic wavefunction, which can be readily handled using the method of Lagrange multiplier. Consequently, the intrinsic motion is still described by the original particle co-ordinates, and hence the intrinsic wavefunction is the shell-model wavefunction but subject to a well-defined constraint. Therefore, the intrinsic wavefunction permits us to use the available and well-known and extensive collection of shell-model concepts and tools.

---

[5] The number in the brackets indicates percentage over-prediction.

From the transformed Bohr-Mottelson Schrodinger equation we derive two self-consistent, coupled equations for the oscillations and intrinsic motion of the nucleus using a mean-field representation of the usual monopole-monopole interaction, a product wavefunction, and a constrained variational method. Each of the two coupled equations resembles that of the conventional phenomenological cranking model and contains its own cranking parameters and energy. These parameters and energy are self-consistently determined by the solutions of the two equations themselves. Consequently, the two equations are time-reversal invariant, unlike the conventional phenomenological cranking-model equation. This time-reversal invariance and self-consistency require real eigenfunctions of the two coupled cranked equations. A sum rule is used to determine the oscillation frequency as a function of the intrinsic-system parameters. A monopole constraint is imposed on the intrinsic wavefunction in its first moment form using Lagrange multiplier and first-order perturbation methods.

Eigenvalues and eigenfunctions of the derived two coupled cranked equations, the corresponding values of the equation parameters, and the excitation energy of the first excited $0^+$ state are determined. Eigenfunction orthogonality and finiteness are shown to impose limits on the range of the values of the cranking parameters.

We have performed a preliminary scoping calculation of the excitation energy of the first excited $0^+$ states in the light nuclei in the range $A \leq 120$ using the mean-field harmonic oscillator potential. Since presently the model uses a somewhat crude approximation to impose the monopole constraint on the intrinsic system, neglects the centre-of-mass motion, the residual two-body interaction (for example, pairing and Nilsson's shell-model corrections), and deformation, this preliminary calculation is not intended to accurately predict this excitation energy but rather to explore the relative importance of the various features, approximations (such as the constraint on the intrinsic system and its first-order treatment), and parameters used in the model. In particular, the residual interaction in any of its (microscopic or phenomenological) forms can have a significant impact on the excitation energy. Therefore, we do not expect and do not obtain a reasonable agreement with the experimental data, although we may have a better success, with the present simple model, in predicting the excitation energy of giant monopole resonances and quadrupole vibration excitation (when nuclear deformation is included) as the corresponding excitation energies may be less sensitive to the neglected factors. In any case, the calculation results indicate that, except for $^4_2He$, $^{12}_6C$ and $^{28}_{14}Si$ where monopole oscillations is the dominant component, the excitation energy of the first excited $0^+$ states for the light nuclei in the range $A \leq 120$ is significantly overpredicted since these nuclei are deformed in their first excited $0^+$ states and hence undergo quadrupole vibration and/or rotational motions.

In a future article, we intend to include in the model: more realistic treatment of and higher moments of the monopole constraint on the intrinsic system, the centre-of-mass motion, the rotational-invariance constraint in Eq. (3), residual two-body interaction such as pairing, and Nilsson shell-model features (i.e., spin-orbit and $l^2$ coupling terms) and examine the impact of these factors on the model predictions. We also intend to include in the model deformation degree of freedom to derive a microscopic self-consistent cranking model for deformed nuclei and reveal the nature of the approximations and assumptions underlying the corresponding conventional phenomenological cranking model, and examine their impact on the predicted rotational-vibrational motion and on the reduced transition rates.